\documentclass[10pt,preprint]{aastex}
\usepackage{times}
\newcommand{\vect}[1]{\ensuremath{\mbox{\boldmath $#1$}}}

\newcommand{\smc}{MACHO~98-SMC-1}
\newcommand{\mb}{MACHO~99-BLG-47}
\newcommand{\eb}{EROS~BLG-2000-5}
\newcommand{\obp}{OGLE-1999-BUL-36}
\newcommand{\obc}{OGLE-1999-BUL-23}

\slugcomment
{to appear in the Astrophysical Journal (v572n2; 2002 June 20)}
\shorttitle{A SHORT, NON-PLANETARY, MICROLENSING ANOMALY}
\shortauthors{the PLANET collaboration}

\begin{document}

\title{A Short, Non-Planetary, Microlensing Anomaly:\\
Observations and Lightcurve Analysis of \mb}

\author{
M. D. Albrow\altaffilmark{1},
J. An\altaffilmark{2},
J.-P. Beaulieu\altaffilmark{3},
J. A. R. Caldwell\altaffilmark{4},
D. L. DePoy\altaffilmark{2},\\
M. Dominik\altaffilmark{5},
B. S. Gaudi\altaffilmark{6,7},
A. Gould\altaffilmark{2},
J. Greenhill\altaffilmark{8},
K. Hill\altaffilmark{8},
S. Kane\altaffilmark{9},\\
R. Martin\altaffilmark{10},
J. Menzies\altaffilmark{4},
R. W. Pogge\altaffilmark{2},
K. R. Pollard\altaffilmark{1},
P. D. Sackett\altaffilmark{5},\\
K. C. Sahu\altaffilmark{11},
P. Vermaak\altaffilmark{4},
R. Watson\altaffilmark{8},
and
A. Williams\altaffilmark{10}
}
\author{(The PLANET Collaboration)}

\altaffiltext{1}
{Department of Physics \& Astronomy, University of Canterbury,
Private Bag 4800, Christchurch, New Zealand}
\altaffiltext{2}
{Department of Astronomy, the Ohio State University,
140 West 18th Avenue, Columbus, OH 43210, USA}
\altaffiltext{3}
{Institut d'Astrophysique de Paris, INSU-CNRS,
98 bis Boulevard Arago, F 75014 Paris, France}
\altaffiltext{4}
{South African Astronomical Observatory,
P.O. Box 9, Observatory, 7935 South Africa}
\altaffiltext{5}
{Kapteyn Institute, Rijksuniversiteit Groningen,
Postbus 800, 9700 AV Groningen, The Netherlands}
\altaffiltext{6}
{School of Natural Sciences, Institute for Advanced Study,
Einstein Drive, Princeton, NJ 08540, U.S.A.}
\altaffiltext{7}
{Hubble Fellow}
\altaffiltext{8}
{Physics Department, University of Tasmania,
G.P.O. 252C, Hobart, Tasmania 7001, Australia}
\altaffiltext{9}
{School of Physics \& Astronomy, University of St.~Andrews,
North Haugh, St.~Andrews, Fife KY16 9SS, UK}
\altaffiltext{10}
{Perth Observatory,
Walnut Road, Bickley, Western Australia 6076, Australia}
\altaffiltext{11}
{Space Telescope Science Institute,
3700 San Martin Drive, Baltimore, MD 21218, USA}

\begin{abstract}

We analyze PLANET and MACHO observations of \mb, the
first nearly-normal microlensing event for which high signal-to-noise-ratio
data reveal a well-covered, short-duration anomaly.
This anomaly occurs near the peak of the event.
Short-duration anomalies near the peak of
otherwise normal events are expected to arise both from
extreme-separation (either very close or very wide),
roughly equal-mass binary lenses, and from planetary systems.
We show that the lens of \mb\ is in fact an extreme-separation binary, 
not a planetary system, thus demonstrating for the first time 
that these two important classes of events can be distinguished in practice.
However, we find that the wide-binary and close-binary lens solutions fit the
data equally well, and cannot be distinguished even at $\Delta\chi^2=1$.
This degeneracy is qualitatively much more severe than the one
identified for \smc\ because the present degeneracy spans two rather
than one dimension in the magnification field and
does not require significantly different blending fractions.
In the appendix, we explore this result, and show that it is
related to the symmetry in the lens equation.

\end{abstract}

\section{Introduction}

	The hallmark of planetary microlensing events is a short
deviation from an otherwise normal, point-source/point-lens (hereafter PSPL)
event.  \citet{mp} showed that extrasolar planets could be detected from 
such events, and \citet{gl} gave an explicit prescription for how the
planet/star mass ratio $q$ ($\ll 1$) and the angular separation $d$ 
(in units of the angular Einstein radius $\theta_{\rm E}$) could
be reconstructed by decomposing the event light curve into its 
``normal'' and ``perturbed'' components.

	Work during the ensuing decade has elucidated many additional
subtleties of planetary light curves, but their fundamental characterization
as briefly perturbed PSPL events has remained intact.   Of particular
importance in the present context, \citet{gs} showed that events with
small impact parameter ($u_0\ll 1$; where $u_0$ is the minimum separation
between the source and the lens center of mass in units of $\theta_{\rm E}$) 
probe the so-called ``central caustic'' of the lens 
geometry, making them much more sensitive to the presence of 
planets than the larger impact-parameter events analyzed by \citet{gl},
which probe the outer ``planetary caustic''.

        These central-caustic events are of exceptional importance,
even though they are intrinsically rare.  They are rare simply
because the central-caustic is much smaller than the planetary caustic,
so the great majority of planet-induced deviations (of fixed
fractional amplitude) are due to planetary caustics.  However, the
probability of detecting a planet is much greater in small impact-parameter
events, partly because the source is guaranteed to pass close to the 
central caustic, which almost by definition is near the center of the
lens geometry ($\vect{u}=0$) and partly because even the sensitivity of 
planetary caustics is enhanced for $u_0\ll 1$.  (Here $\vect{u}$ denotes the
source position on the sky, normalized to $\theta_{\rm E}$,
with respect to the lens center of mass.)  By contrast, higher 
impact-parameter events miss the central caustic, and they are likely to miss
the planetary caustic as well because it lies in a random position
relative to the source trajectory.  Because of their higher sensitivity
to planets, and because they can be recognized in real time,  low
impact-parameter events are monitored more intensively than typical events
by microlensing follow-up networks, which in turn further enhances
their sensitivity to planets.

Central-caustic
events, like their planetary-caustic cousins, involve a short deviation
from an otherwise normal PSPL light curve.  
The major difference between these two classes of planetary
events is that central-caustic anomalies always occur near the peak,
whereas planetary-caustic perturbations can occur
anywhere on the light curve, and are typically expected on the
wings of the light curve.
Of particular importance, for
central-caustic events, there is
no simple prescription for extracting $d$ and $q$ by decomposing the
light curve into ``normal'' and ``perturbed'' components and it is
unclear to what degree these two parameters are degenerate.

Another important, albeit accidental, discovery was that planets could
give rise to perturbations that are not short compared to the event
timescale.  In the course of their search for planetary perturbations
among 43 approximately PSPL events, 
\citet[see also \citealt{letter}]{planetsearch}
found one event, \obp,
that was asymmetric in a way that was consistent with the presence of
a planet.  They argued, however, that the asymmetry was also consistent
with parallax effects induced by the Earth's motion around the Sun of the
type analyzed by \citet*{gmb}, and that in general it would be extremely 
difficult to distinguish between the two possible causes of such an
asymmetry. They concluded that, in most cases, microlensing
searches are not able to distinguish between parallax and
a weak, asymmetric planetary perturbation,
and consequently, all such ``detections'' should be ignored.
This reduces the sensitivity of microlensing searches to planets,
but only by an extremely small amount since, as \citet{planetsearch}
showed, long-timescale asymmetric perturbations account for 
less than $\sim 1\%$ of all planetary events. Hence, the long-timescale
asymmetric events also confirm in a way the basic paradigm:
planetary perturbations have short durations relative to the
parent light curve, and in the rare cases for
which they do not, they are not recognizable as a planetary perturbation
anyway.

However, not all short timescale deviations are due to planets and
therefore
the mere detection of such an anomaly does not prove the presence of a
planet. \citet{gg} showed that close binaries ($d\ll 1$) give rise to
light curves that are virtually identical to PSPL events, except when
the source comes very close to the lens center of mass ($|\vect{u}|\ll 1$).
Hence,
for events with $u_0\ll 1$, the light curve looks ``normal'' except for a
brief deviation near the peak.  Qualitatively, this is exactly the
same behavior as that of central-caustic planetary events.  
Similarly, light curves of wide-binary ($d\gg 1$) events 
can also take the same form if one -- and only one -- of the caustics lies
very close to the source's passage. Indeed,
a close correspondence between a certain pair of
close-binary and wide-binary events was discovered both
theoretically \citep{d99} and observationally \citep{comp,smc}.
It remains an open question
under what conditions these various types of events can be distinguished
from one another.  If central-caustic planetary events could not be 
distinguished from close- and/or wide-binary events,
it would seriously undermine 
planet searches in high-magnification events and hence would call into
question the basic strategy adopted by microlensing follow-up groups
\citep[e.g.,][]{PLANET}.

Here we analyze the light curve of the microlensing event \mb,
the first intensively monitored event with
a short-lived, high signal-to-noise-ratio deviation
from an otherwise normal PSPL light curve.   
We identify the lens as an extreme-separation binary rather than a 
planetary system, thereby showing that,
at least in this case, the two classes can be clearly distinguished.
We also show that both wide- and close-binary solutions fit the
data equally well, implying that although planetary perturbations can be
distinguished from those arising from extreme-separation binaries,
the discrimination between very close and very wide binaries may
be difficult in practice.

\section{\mb}

The initial alert for the microlensing event \mb\ was issued
by the MACHO collaboration on 1999 July 23.\footnote{
\anchor{ftp://darkstar.astro.washington.edu/macho/Alert/99-BLG-47/}
{\url{ftp://darkstar.astro.washington.edu/macho/Alert/99-BLG-47/}}}
The event was located about 18\degr\ from the Galactic center along
the disk 
($l = 17\arcdeg 59\arcmin 10\arcsec$, $b= -1\arcdeg 57\arcmin 36\arcsec$)
and reported to be rather faint at baseline ($V=21.5$, $R=20.3$).
PLANET began monitoring \mb\ right after
the electronic alert with the expectation that it would be a very high
magnification event. PLANET detected anomalous behavior in the light curve 
during the first week of August, and subsequently issued
an anomaly alert on 1999 August 4.\footnote{
\anchor{http://www.astro.rug.nl/~planet/MB99047.html}
{\url{http://www.astro.rug.nl/\~{}planet/MB99047.html}}}

\subsection{Observations and Data
\label{sec:obs}}

The PLANET light curve of \mb\ consists of
observations made from three different southern sites:
the Elizabeth 1 m at the South African Astronomical Observatory (SAAO),
Sutherland, South Africa; 
the Canopus 1 m near Hobart, Tasmania, Australia;
and the Yale/AURA/Lisbon/OSU (YALO) 1 m at Cerro Tololo Inter-American
Observatory (CTIO), La Serena, Chile. The observations were carried out
in two bands: $I$ (at all three sites) and $V$ (at SAAO only).
The event was most intensely monitored in 1999 August
during and following the photometric peak and the anomaly, but we
obtained data through mid-September and also when the event was
close to baseline early in the 2000 season as the source came out from 
behind the Sun.  After the usual reductions, we
perform photometry by two independent methods: a direct fit
to the point spread function (PSF) using DoPHOT \citep*{dophot} and difference
imaging using ISIS \citep{isis1,isis2}.  For details, 
see also \citet{mb41}.
Because of the different characteristics of the two methods
and varying frame qualities, we recover a
different number of photometric measurements. Before any elaborate
efforts to clean and select the data, we have 303 points from DoPHOT
photometry and 297 from ISIS photometry (Table~\ref{tab:data}) for the 
1999 observations, as well as 8 points observed from YALO during
early 2000 season (DoPHOT photometry only).

The photometric observations taken at SAAO are calibrated to the standard
Johnson-Cousins system with respect to E-region stars \citep{calib}
observed contemporaneously with \mb\ at SAAO on 1999 July 31. 
The instrumental magnitudes of observations from Canopus and YALO
are not calibrated explicitly, but rather we allow independent source
fluxes for different sites and bands in our modeling,
which provide the relative photometric scalings among them.

We also include in our analysis the publicly available MACHO photometry,
which comprises 77 $V_{\rm MACHO}$ and 90 $R_{\rm MACHO}$ points
from the 1999 season
as well as 213 $V_{\rm MACHO}$ and 227 $R_{\rm MACHO}$ points taken
between 1995 and 1998.  The latter were used to constrain the baseline 
and to check for source variability. We find no evidence for such
variability.

The difference photometry is placed on an absolute scale by
a linear regression from the PSF-based flux (DoPHOT) to the differential
flux (ISIS), after removing obvious faulty points 
from the two data sets based on the reports by each reduction/photometry
package.  We find that, except for the Canopus data, the residuals
of the PSF-based flux from the regression line
are strongly correlated with the size of the seeing disk,
but that most of this correlation is removed by adding a linear seeing 
correction. For example, for the SAAO $I$-band observations,
we detect the presence
of a non-zero linear seeing dependence term of
$\sim 1.7 F_{\rm base}\ \mbox{arcsec}^{-1}$
(or $\sim 10 F_{\rm s}\ \mbox{arcsec}^{-1}$),
with the signal to noise ratio being as large as $\sim 30$.
Here $F_{\rm s}$ and $F_{\rm base}$ are
the net flux of the lensed source star alone and
the total flux in the same PSF when there is no magnification.

In order to minimize the effects of known systematics, we optimize the
data set by rejecting outliers and reevaluating
the sizes of the photometric error bars.
We also include a correction term for the correlation between
the seeing and the blended flux that enters
the same PSF with the lensed source
when we fit the observed flux to a specific magnification model.
These procedures are described in detail and fully justified in several
earlier PLANET papers \citep{mb41,ob14,ob23,letter,planetsearch}.  
Briefly, we first construct
an initial clean subset of the data by rejecting 3-$\sigma$ outliers with
respect to the linear regression from DoPHOT to ISIS flux,
and then determine a reference model
based on this subset. Once we have a reference model, we include all the
data and follow the same iterative procedure of outlier removal and
error rescaling described in \citet{ob23}.
The final data set used for the subsequent analysis reported in
the present paper contains 276 DoPHOT-reduced and 266 ISIS-reduced PLANET
points as well as 166 MACHO points (one $V_{\rm MACHO}$ point rejected),
all from the 1999 season (Table~\ref{tab:data}).
The data obtained during seasons
other than 1999 (the 2000 season YALO $I$, and the 1995 -- 1998 season
MACHO $V_{\rm MACHO}$ and $R_{\rm MACHO}$ bands) are included in
the analysis as a combined single point for each set.
In doing so, 4 data points (all $R_{\rm MACHO}$ in the 1998 season)
are eliminated as outliers.

\section{Analysis and Model
\label{sec:model}}

\begin{figure}
\plotone{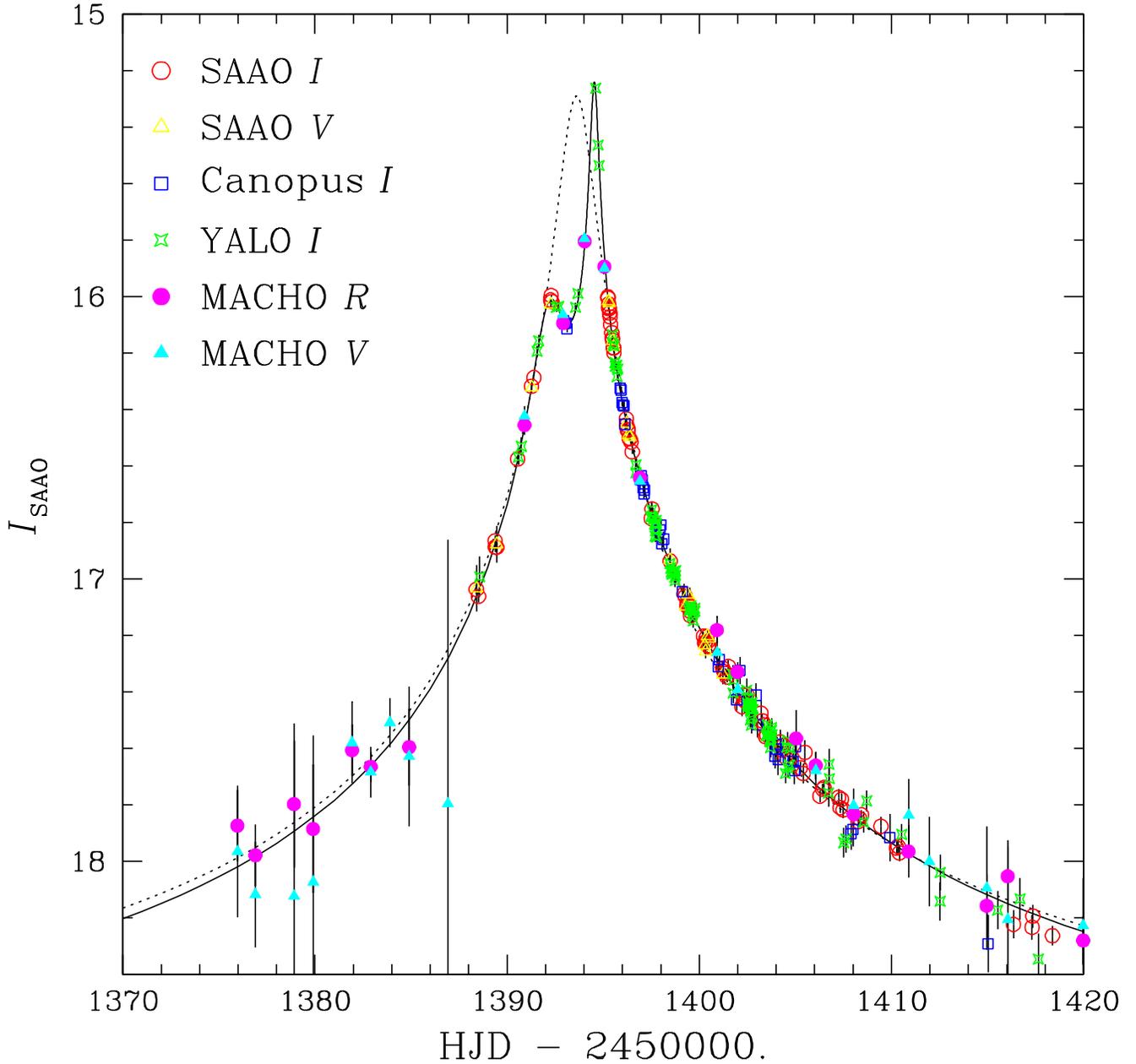}
\figcaption{
Observations and models of \mb, all scaled to calibrated Cousins $I$-band
with blending as registered by SAAO DoPHOT photometry.  Shown are
PLANET $I$-band from
SAAO, Canopus, YALO,
PLANET $V$-band from SAAO, 
and MACHO $V_{\rm MACHO}$ and $R_{\rm MACHO}$.
The ISIS reduced PLANET points are plotted after applying
the transformation to the absolute scale derived from
the linear regression between difference photometry and PSF-based photometry.
The solid curve shows the final 
close-binary lens model fit (Table~\ref{tab:model}) to the data
while the dotted curve shows the ``best'' degenerate form of
point-source/point-lens light curve (eq.~[\ref{eqn:degen}])
fit to a high-magnification subset of the data
that excludes the anomalous points near the
peak.  The half-magnitude offset between this curve and the data
is the main observational input into the algorithm to search for the
final model (see \S~\ref{sec:model}).  Note that on the scale
of the figure, the wide-binary solution is completely
indistinguishable from the close-binary solution, i.e. the solid
curve can represent both the close-binary and the wide-binary lens models
equally well.
\label{fig:data}
}\end{figure}

	The light curve of \mb\ (Fig.~\ref{fig:data}) is that
of a normal high-magnification
PSPL event with a short-lived ($\sim 3\ \mbox{days}$)
deviation near the peak, of the type predicted for planets \citep{gs}
or binaries with extreme separations. We therefore
develop a method for probing the space of lens geometries of this type
and then search for the minimum $\chi^2$ within this space.

	We begin by excising the anomalous points near the peak and
fitting 
the remaining light curve to the degenerate form of the \citet{p86}
profile that obtains in the high-magnification limit \citep{g96},
\begin{equation}
\label{eqn:degen}
F(t) = 
\frac{F_{\rm peak}}{\sqrt{1 + (t-t_0)^2/t_{\rm eff}^2}} + F_{\rm base}
\ .\end{equation}
Here $F(t)$ is the instantaneous flux,
$F_{\rm peak}$ is the peak flux above the baseline of the model,
$F_{\rm base}$ is the baseline flux,
$t_0$ is the time of the peak,
and $t_{\rm eff}$ is the effective width of the light curve.
These degenerate parameters are related to the standard
parameters by $F_{\rm peak} = F_{\rm s}/u_0$, 
$F_{\rm base} = F_{\rm s} + F_{\rm b}$, and 
$t_{\rm eff} = u_0 t_{\rm E}$,
where $t_{\rm E}$ is the Einstein timescale, i.e., the time required for
the source to move an Einstein radius.
We find $t_0\sim 2451393.6$ (Heliocentric Julian Date),
$t_{\rm eff}\sim 1\ \mbox{day}$, and that $F_{\rm peak}$ 
roughly corresponds to $I_{\rm peak} \sim 15.3$ (Fig.~[\ref{fig:data}]).
However, the exact values of those parameters are quite dependent on
which points we excise. Note that though the baseline magnitude
is measured to be $I_{\rm base} = 19.1$, there is at this point
essentially no information of how the baseline flux divides into
source flux $F_{\rm s}$ and blended flux $F_{\rm b}$,
which is why it is 
necessary to fit the light curve to the degenerate profile 
(eq.~[\ref{eqn:degen}]).

	Next, we reinsert the excised points near the peak and note
that the maximum (over-magnified) deviation from the degenerate fit 
is $\sim 0.5\ \mbox{mag}$ (Fig.~\ref{fig:data}).
We then establish a set of initial trial event geometries as follows.
For each diamond-shaped caustic produced by various geometries of
($d$,$q$) pairs, we examine the magnification as a function of
distance from the ``caustic center'' along each of the three directions 
defined by the four cusps of the caustic
(one direction is redundant due to
the reflection symmetry with respect to the binary axis).
Here for computational simplicity, we use an analytic proxy point for the
``caustic center,'' defined as follows.
For close binaries, we adopt the center of mass of the
lens system for the caustic center. On the other hand, for wide binaries,
the position of one component of the binary shifted toward the other
component by $[d(1+q^{-1})]^{-1}$ is chosen as the caustic center.
[These choices can be understood from the approximation developed
in Appendix~\ref{sec:ap}. See also \citet{d99}.]
Then, at each point $\vect{u}_\bullet$, the point of the cusp-axis
crossing, we take the ratio of
the actual magnification to that predicted for the corresponding
PSPL lens: a point mass lens located at the caustic center
with the same mass as either the whole binary (close binaries) or
the nearest lens alone (wide binaries). We proceed with the search
only if this ratio lies in the interval $[1.2,1.8]$. We then choose
a source trajectory perpendicular to the cusp axis
as an initial trial model, with $u_0 = |\vect{u}_\bullet|$ and 
$t_{\rm E} = t_{\rm eff}/|\vect{u}_\bullet|$, and
use four-parameter ($\alpha$, $t_{\rm E}$, $t_0$, and $u_0$)
downhill simplex \citep{nr} to search for a best-fit model
to minimize $\chi^2$, which is determined
by a linear-flux fit to the model of lens magnifications (and thus the
corresponding $F_{\rm s}$ and $F_{\rm b}$ are found immediately),
subject to the constraint of fixed ($d$,$q$).
Figure~\ref{fig:chi2cont} shows the result of this search
(based on ISIS solutions)
in a contour plot of $\chi^2$-surface over the ($d$,$q$)-space.
The $\chi^2$ is rising around all the boundaries
shown in Figure~\ref{fig:chi2cont}, except toward lower $q$.
While the $\chi^2$-surface flattens with $\Delta\chi^2\simeq 32$
as the mass ratio $q$ becomes very small ($q\la 0.01$)
for $d\sim 0.065$ and $d\sim15$, we find no evidence of a
decrease of the $\chi^2$ as $q$ becomes smaller than $0.01$.
Rather it asymptotically approaches $\Delta\chi^2\simeq 32$.

We find well-localized minima of $\chi^2$ over 
the searched ($d$,$q$)-space, one of close binaries ($d\simeq 0.13$) and
the other of wide binaries ($d\simeq 11.3$), whose exact parameters
depend slightly on whether we use the ISIS or DoPHOT photometry
(Tables~\ref{tab:model} and \ref{tab:modelw}).  
We adopt the ISIS solutions in the subsequent discussion
(and they are also what is shown in Fig.~\ref{fig:chi2cont})
because the light curve shows less scatter and consequently,
the errors for the model parameter determinations are smaller.
Despite the combination of the dense coverage near the peak by the
PLANET data and the extensive baseline coverage by the MACHO data,
the final two ISIS models (i.e., the close binary and the wide binary)
are essentially indistinguishable: $\Delta\chi^2=0.6$ for 412 degrees
of freedom, with the close binary having the lower $\chi^2$ of the two.
In Appendix~\ref{sec:ap}, we discuss this degeneracy in further detail.

	Finally, we also note that $u_0 t_{\rm E}$ and $F_{\rm s}/u_0$
from the best-fit binary-lens model parameters are not quite the same as
the $t_{\rm eff}$ and $F_{\rm peak}$ derived from the initial degenerate
form of the light curve fit. These discrepancies are traceable to small
differences between a true \citet{p86} curve and its degenerate
approximation (eq.~[\ref{eqn:degen}]), and are not due to differences
between the binary and corresponding PSPL event, which are very small
except around the peak. Since the parameters derived from
equation~(\ref{eqn:degen}) function only as seeds for simplex, and
since the final $\chi^2$-surface is very well-behaved,
these discrepancies in input values do not influence the final result.

\section{Discussion}

	The mass ratios of the best fit models are $q=0.340\pm 0.041$
(close binary) and $q=0.751\pm 0.193$ (wide binary),
which are well away from the regime of planetary companions 
($q\la 0.01$). By comparison,
the ratio of the duration of the anomalous portion of the light curve
($t_{\rm anom}\sim 3\ \mbox{days}$) to the Einstein timescale
($t_{\rm E}\sim 160\ \mbox{days}$)\footnote{While the value here is for
the close binary, the value for the wide binary is essentially the same
because the relevant mass of the corresponding point-mass lens for the
wide binary is not the combined mass but
the mass of the single component that the source passes by.}
is $r=t_{\rm anom}/t_{\rm E} \sim 0.02$.
For light curves perturbed by planetary
caustics, one typically finds $q\sim r^2$ \citep{gl}.  This relation
clearly does not apply to the caustics of extreme-separation binaries.  

	Models with $q\le 0.01$ (and $q\ge 100$) are formally
rejected at $\Delta\chi^2\simeq 32$, which is significant but
not in itself an overwhelming rejection of the planetary
hypothesis. Hence, we
examine the least $\chi^2$ model for $q=0.01$ ($\Delta\chi^2=32.3$)
for its plausibility and the origin of statistical discriminating
power.  We discover that the difference of $\chi^2$ is mostly
from the MACHO data between HJD $2451310$ and $2451360$,
approximately 2 months prior to the photometric peak.
However, we also find that the ``planetary'' model
exhibits an extremely high peak magnification
($A_{\rm max}\sim 15000$), and consequently, requires the event
to be much longer ($t_{\rm E}\simeq 40\ \mbox{yr}$)
and more extremely blended ($I_{\rm s}\simeq 25.7$,
$F_{\rm s}/F_{\rm base}\simeq 0.2\%$) than the already unusually
long and highly blended best-fit models. In addition, there is a clear
trend of increasing peak magnification (and thus timescale and blend
as well) as $q$ is lowered beyond $0.01$. This follows from the fact
that the observed timescale of the anomaly in the light curve essentially
fixes the source movement relative to the caustic. However, as $q$
becomes smaller, the size of the caustic relative to the Einstein ring
shrinks, and therefore, the timescale of the event, that is the time required
for the source to cross the Einstein ring, increases.
(Note that this behavior causes a mild continuous degeneracy
between $q$ and the blending.)
These parameters determined for the ``planetary'' model are
extremely contrived and highly improbable a priori.
Furthermore, the timescale associated with the ``planet'' component,
$t_{\rm p}=q^{1/2}t_{\rm E}\sim 4\ \mbox{yr}$ is much longer than that
of typical stellar lenses, which further reduces the plausibility
of the planetary interpretation.
In summary, while in simple statistical terms, 
the star/planet scenario is not overwhelmingly disfavored relative to
extreme-separation binaries,
we nevertheless can conclude that it is highly unlikely that
this event is due to a star/planet system.

        From the standpoint of refining microlensing planet detection
strategies, it is important to ask how one could have discriminated
between the planetary and extreme-separation binary solutions with greater
statistical significance.  As noted above, most of the discriminating
power came from the MACHO data points on the rising wing of the light
curve, even though (or in a sense, because) these had the largest
errors and the lowest density of the non-baseline coverage.  That is,
the precision PLANET photometry over the peak and falling wing
``predicts'' the rising wing for each of the models, but the noisier
MACHO data can only roughly discriminate between these predictions.
Hence, the key would have been to get better data on the rising part
of the light curve.  In practice this is difficult: the MACHO data
are noisier exactly because they are survey data, and one does not
know to monitor an event intensively until the light curve has actually 
started to rise.

	The long duration of the event, $t_{\rm E}\simeq 160\ \mbox{days}$
(close binary) or $t_{\rm E}\simeq 220\ \mbox{days}$ (wide binary), may
lead one to expect that the event would show some sign of parallax effects
\citep{g92,m99,s01,b01}.
Similarly, the close passage of the source to a cusp
could in principle give rise to finite source effects
\citep{g94,nw,wm}, as for example
was recently observed in the case of \eb\ \citep{eb05}.
Indeed, the combination of parallax and finite-source effects permitted
\citet{eb05} to measure the mass of a microlens for the first time.
We have therefore searched for both parallax and finite source effects,
but find no significant detection of either.

	Finally, we examine the position of the source on the
color-magnitude diagram (CMD; Fig.~\ref{fig:cmd}). 
The most prominent feature found in the CMD is a track of stars running
diagonally in the same direction as would the main sequence.  Since the
field is in the Galactic disk, this feature is probably not a true
main sequence, but rather likely to be a ``reddening sequence'', that is,
an ensemble of mostly disk turnoff stars at progressively greater
distances and correspondingly greater reddenings.
The baseline ``star'' (combined light of the source and blended light)
lies within this sequence toward its faint/red end although
its position is seeing-dependent because of the seeing-correction
term. The position indicated in the CMD is plotted assuming median seeing.
However, the source itself [$I_{\rm s}=20.9$, $(V-I)_{\rm s}=1.94$]
lies $\sim 2\ \mbox{mag}$
below the baseline ``star,'' in a region of the CMD that is well below the
threshold of detection.
If the baseline were mainly composed of source light, and
the ``blended light'' were simply source light that had been falsely
attributed to blending by a wrong model, then the real source would lie
within the well populated ``reddening sequence'',
and the timescale would be much shorter, $t_{\rm E}\sim 30\ \mbox{days}$.
We therefore searched for solutions with little
or no blending.  However, we find that these are ruled out with
$\Delta\chi^2=2440$.  Furthermore, the best-fit models with no blended light
are still extreme-separation binaries and not planetary systems. 
Since we have only a crude
understanding of the CMD, no strong argument can be made that the
position of the source is implausible. If, in fact, the observed
track of stars in Figure~\ref{fig:cmd} is really the reddening sequence
of disk turnoff stars, the source position is consistent with
a low-mass main-sequence star lying $\ga 2\ \mbox{mag}$
below the turnoff. This would also explain the lack of
finite source effect.
Moreover, we note that in disk fields, long events are not
at all uncommon \citep{derue} because the observer, source, and lens
are all moving with roughly the same velocity.

Regarding the large amount of blended light required to fit
the observed light curve, we note that the highly significant
seeing effect in the PSF-based photometry 
(see \S~\ref{sec:obs}) is also evidence
that the event is strongly blended. In addition to the direct
confirmation of the strong seeing effect from the comparison 
between the differential flux and the PSF-based flux measurements,
we independently detect a significant seeing correction when
we fit the observed flux to the magnification model.
For SAAO $I$-band observation, the seeing correction determined
from the model fit using DoPHOT flux is slightly smaller
($\sim 9 F_{\rm s}\ \mbox{arcsec}^{-1}$) than the value 
derived from the regression between DoPHOT and ISIS flux.
The seeing correction derived by fitting ISIS data to the 
model is basically consistent with zero.

\section{Binary Lens vs.\ Binary Source}

Multiple-peak events like the one seen in Figure~\ref{fig:data}
can be caused by a binary {\it source} \citep{gh}
rather than a binary {\it lens}.  In general, one expects that
such events will be chromatic, since the colors of the two sources
will not usually be the same.  By contrast, \mb\ is
achromatic: the difference in color of the two wings of the light curve 
is $\Delta V-I = 0.01\pm 0.05$.  In the limit that one component 
of the binary is completely dark, the event will be achromatic, and
yet can have still have multiple peaks caused by rotation of the binary 
during the event.  In this case, however, there will be a series of
roughly equally spaced peaks that gradually die out as the event
declines \citep{hg}, contrary to the distinct double-peaked
behavior seen in Figure~\ref{fig:data}.  Hence, a binary-source explanation
appears a priori implausible.

        Nevertheless, it is possible in principle that the source
has two components of nearly identical color.  We search for binary-source
models, with either static or slowly moving components, but find that the
best fit model has $\Delta\chi^2=1167$, which clearly rules out
such models.  The basic problem is that the observed second peak is
very sharp given its height and the timescale of the declining light curve.
Binary-source light curves that fit these latter features tend to have
a width that is closer to that of the dotted curve in Figure~\ref{fig:data}.
Recall that this curve represents a single-lens degenerate fit to the non-peak
data. We conclude that the explanation for the double-peaked behavior of the
light curve is that it is a binary-lens rather than  a binary-source event.

\section{Conclusion}

	\mb\ is the first microlensing event with a short-lived,
high signal-to-noise-ratio anomaly, characteristics that could betray
the existence of a planet around the lensing star.
Nevertheless, we conclude that
the lens of \mb\ is not a planetary system,
but an extreme-separation (very close or very wide) binary
composed of components of similar mass, based on the result of the
light curve fit as well as the extreme value of event duration and
blending fraction required for any plausible ``planetary'' fit.

\acknowledgements
\medskip{\center \bf ACKNOWLEDGEMENTS}\\
We thank the MACHO collaboration for providing the initial alert for
this event and for allowing us to use their photometric data for the
modeling.
We thank Joachim Wambsganss for his comments on and a thorough
reading of the draft version of this paper.
We are especially grateful to the observatories that support our science
(Canopus, CTIO, SAAO) via the generous allocations of time
that make this work possible. The operation of Canopus Observatory
is in part supported by the financial contribution from
Mr.\ David Warren.
PLANET acknowledges financial support via
award GBE~614-21-009
from de Nederlandse Organisatie voor Wetenschappelijk Onderzoek (NWO),
the Marie Curie Fellowship ERBFMBICT972457
from the European Union (EU),
``coup de pouce 1999'' award
from le Minist\`ere de l'\'Education nationale,
de la Recherche et de la Technologie,
D\'epartement Terre-Univers-Environnement,
the Presidential Fellowship
from the Graduate School of the Ohio State University,
a Hubble Fellowship 
from the Space Telescope Science Institute (STScI),
which is operated by
the Association of Universities for Research in Astronomy (AURA), Inc.,
under NASA contract NAS5-26555,
the Jet Propulsion Laboratory (JPL) contract 1226901,
grants AST~97-27520 and AST~95-30619
from the National Science Foundation (NSF),
and grants NAG5-7589 and NAG5-10678
from the National Aeronautics and Space Administration (NASA).
This paper utilizes public domain data obtained by the MACHO Project,
jointly funded by the US Department of Energy through the University
of California, Lawrence Livermore National Laboratory under contract
No.\ W-7405-Eng-48, by the NSF through
the Center for Particle Astrophysics of the University of California
under cooperative agreement AST-8809616, and by the Mount Stromlo and
Siding Spring Observatory, part of the Australian National University.

\appendix
\section{Close and Wide Binary Correspondence
\label{sec:ap}}
\citet{d99} noticed that the \citet{crn,cr} lens 
\citep*[CRL; see also][]{sef}
well approximates the 
binary lens system in the vicinity of each individual
lens component of extreme wide-separation ($d\gg 1$) binaries and
in the vicinity of the secondary of extreme close-separation ($d\ll 1$)
binaries while the quadrupole lens (QL) approximation works nicely near the
center of mass of extreme close-separation binaries. He further
showed that the size and shape of the caustics and the behavior of the
magnification field associated with them are very similar among
the true binary lenses, the CRL approximation, and the QL approximation
if the shear, $\gamma$, of the CRL and 
the absolute eigenvalue of the quadrupole moment tensor, $\hat Q$
(for simplicity, hereafter $\hat Q$ will be just
referred to ``the quadrupole moment''),
of the QL are both very small and their numerical values are close
to each other ($\gamma\simeq\hat Q\ll 1$).
These findings also imply the existence of a correspondence between
a wide binary with $\gamma=d_w^{-2} q_w (1+q_w)^{-1}$ and
a close binary with $\hat Q=d_c^2 q_c (1+q_c)^{-2}$,
sometimes referred to as a $d\leftrightarrow d^{-1}$ correspondence.
(Here and throughout this appendix, we use the subscript $_c$ and $_w$
to distinguish the parameters associated with the close and wide binaries.)
Here we rederive the basic result for this correspondence,
and explore it more thoroughly.

The general form of the binary lens mapping equation can be expressed in
notation utilizing complex numbers \citep{w90},
\begin{equation}
\label{eqn:lenseq}
\zeta = z -
\frac{\epsilon_1}{\bar z - \bar z_1} -
\frac{\epsilon_2}{\bar z - \bar z_2}
\,,\end{equation}
and the magnification associated with a single image can be found by,
\begin{equation}
A = \left(1-\left|\frac{\partial\zeta}{\partial\bar z}\right|^2\right)^{-1}
\,,\end{equation}
where $\zeta$, $z$, $z_1$, and $z_2$ are the positions of 
the source, the image, and the two lens components normalized
by the Einstein radius of some mass, and $\epsilon_1$
and $\epsilon_2$ are the masses of lens components in units of
the same mass.
For the extreme close-binary case, if one chooses the center of mass
as the origin and the binary axis as the real axis
and sets the combined mass to be unity, then $z_1=d_c\epsilon_2$,
$z_2=-d_c\epsilon_1$, $\epsilon_1=(1+q_c)^{-1}$,
and $\epsilon_2=q_c(1+q_c)^{-1}$,
and the lens equation~(\ref{eqn:lenseq}) becomes
\begin{eqnarray}
\zeta&=&z- 
\frac{\epsilon_1}{\bar z - d_c\epsilon_2} -
\frac{\epsilon_2}{\bar z + d_c\epsilon_1}\\\label{eqn:quad}
&\approx&z - \frac{1}{\bar z} -
\frac{d_c^2\epsilon_1\epsilon_2}{\bar z^3} +
\frac{d_c^3\epsilon_1\epsilon_2(\epsilon_1-\epsilon_2)}{\bar z^4} + \cdots
\ \ \ \ (d_c\ll|z|)
\ .\end{eqnarray}
Here the monopole term ($\bar z^{-1}$) is the same as for a PSPL and
the first non-PSPL term is the quadrupole ($\bar z^{-3}$), which
acts as the main perturbation to the PSPL if the quadrupole moment is
small; $\hat Q=d_c^2\epsilon_1\epsilon_2=d_c^2q_c(1+q_c)^{-2}\ll 1$.
For a given source position $\zeta$, let the image position for
the PSPL be $z_0$. Then, under the perturbative approach, the image
for the QL approximation is found at $z=z_0+\delta z_c$, where
$|\delta z_c|\ll|z_0|\sim 1$. Using the fact that $z_0$ is the image
position corresponding to $\zeta$ for the PSPL 
(i.e., $\zeta=z_0-\bar z_0^{-1}$),
one obtains
\begin{equation}
\label{eqn:delc}
\delta z_c\approx\hat Q
\left(1-\frac{1}{|z_0|^4}\right)^{-1}
\left[
\left(\frac{1}{\bar z_0^3}-\frac{1}{z_0^3\bar z_0^2}\right)
+\left(\frac{1}{z_0^4\bar z_0^2}-\frac{1}{\bar z_0^4}\right)
\frac{1-q_c}{1+q_c}d_c+\cdots\right]
\,,\end{equation}
from equation~(\ref{eqn:quad}) and taking only terms linear in
$\delta z_c$ and $\hat Q$. Here, we note that 
$\hat Q\sim\mathcal{O}(d_c^2)$ so that the second term in the bracket
is lower order than $\hat Q^2$.
In order to find the magnification for this
image, we differentiate equation~(\ref{eqn:quad}), and find,
\begin{equation}
\label{eqn:part}
\frac{\partial\zeta}{\partial\bar z}\approx
\frac{1}{\bar z^2} + \frac{3\hat Q}{\bar z^4} 
\left[1-\frac{4(1-q_c)}{3(1+q_c)}\frac{d_c}{\bar z}
+\cdots\right]
\ .\end{equation}
Then substituting $z=z_0+\delta z_c$, for which $\delta z_c$ is given
by equation~(\ref{eqn:delc}), into equation~(\ref{eqn:part}),
we obtain,
\begin{equation}
\left|\frac{\partial\zeta}{\partial\bar z}\right|^2\approx
\frac{1}{|z_0|^4}+\hat Q
\left[
\frac{3|z_0|^4-2|z_0|^2-1}{|z_0|^8-|z_0|^4}
\left(\frac{1}{z_0^2}+\frac{1}{\bar z_0^2}\right)
-\frac{4|z_0|^4-2|z_0|^2-2}{|z_0|^8-|z_0|^4}
\left(\frac{1}{z_0^3}+\frac{1}{\bar z_0^3}\right)
\frac{1-q_c}{1+q_c}d_c+\cdots\right]
\ .\end{equation}
Note that the total magnification for the given source position is
usually dominated by one or two images found close to the critical
curve. Thus, we consider only the case for which the non-perturbed PSPL
images lie close to the unit circle, so we have $|z_0|=1+\Delta$
and $\Delta\ll 1$. Then, we find the expression for the inverse
magnification for the QL approximation (up to the order of $d_c^3$),
\begin{eqnarray}
\label{eqn:magc}
A^{-1}&\approx&
\left|
4\Delta-2\hat Q\left(\frac{1}{z_0^2}+\frac{1}{\bar z_0^2}\right)
+3\hat Q\left(\frac{1}{z_0^3}+\frac{1}{\bar z_0^3}\right)
\frac{1-q_c}{1+q_c}d_c\right|\nonumber\\
&=&4\left|(|z_0|-1)-\hat Q\Re(z_0^{-2})
+\frac{3(1-q_c)}{2(1+q_c)}d_c\hat Q\Re(z_0^{-3})\right|
\ .\end{eqnarray}

For the extreme wide-binary case, one can rewrite
the lens equation~(\ref{eqn:lenseq}) as
\begin{eqnarray}
\zeta&=&z - \frac{1}{\bar z} -
\frac{q_w}{\bar z + d_1}\,,\\\label{eqn:crl}
&\approx&z - \frac{1}{\bar z} - \frac{q_w}{d_1} +
\frac{q_w}{d_1^2}\bar z - \frac{q_w}{d_1^3}\bar z^2 + \cdots
\ \ \ \ (d_w\gg|z|)
\ .\end{eqnarray}
Here the position and the mass of the first lens component are the origin and
the unit mass so that $z_1=0$, $z_2=-d_1$, $\epsilon_1=1$, $\epsilon_2=q_w$,
and $d_1=(1+q_w)^{1/2}d_w$. We note that, apart from the constant
translation, the first non-PSPL term here is
essentially the shear, $\gamma=q_w d_1^{-2}=q_w d_w^{-2}(1+q_w)^{-1}$
for the CRL approximation. Analogous to the extreme close binary, if
$\gamma\ll 1$, the image position for the CRL approximation can be found
by the perturbative approach, but here the corresponding PSPL source 
position would be $\zeta+q_w d_w^{-1}=z_0-\bar z_0^{-1}$. Then, 
the image deviation $\delta z_w=z-z_0$ of CRL from PSPL is
\begin{equation}
\label{eqn:delw}
\delta z_w\approx\gamma
\left(1-\frac{1}{|z_0|^4}\right)^{-1}
\left[
\left(\frac{z_0}{\bar z_0^2}-\bar z_0\right)
+\left(\bar z_0^2-\frac{z_0^2}{\bar z_0^2}\right)
\frac{1}{(1+q_w)^{1/2}d_w}+\cdots\right]
\ .\end{equation}
Using this result and the derivative of equation~(\ref{eqn:crl}),
\begin{equation}
\frac{\partial\zeta}{\partial\bar z}\approx
\frac{1}{\bar z^2} + \gamma
\left[1 - \frac{2}{(1+q_w)^{1/2}}\frac{\bar z}{d_c}
+ \cdots
\right]
\,,\end{equation}
we find
\begin{equation}
\left|\frac{\partial\zeta}{\partial\bar z}\right|^2\approx
\frac{1}{|z_0|^4}+\gamma
\left[
\frac{|z_0|^4+2|z_0|^2-3}{|z_0|^4-1}
\left(\frac{1}{z_0^2}+\frac{1}{\bar z_0^2}\right)
-\frac{2|z_0|^6+2|z_0|^4-4|z_0|^2}{|z_0|^4-1}
\left(\frac{1}{z_0^3}+\frac{1}{\bar z_0^3}\right)
\frac{1}{(1+q_w)^{1/2}d_w}+\cdot
\right]
\ .\end{equation}
From the same argument used for the QL approximation of the
extreme close binary, we can set $|z_0|=1+\Delta$, and then
the inverse magnification for the CRL approximation (up to the
order of $d_w^{-3}$) is
\begin{eqnarray}
\label{eqn:magw}
A^{-1}&\approx&
\left|4\Delta-2\gamma
\left(\frac{1}{z_0^2}+\frac{1}{\bar z_0^2}\right)
+3\gamma\left(\frac{1}{z_0^3}+\frac{1}{\bar z_0^3}\right)
\frac{1}{(1+q_w)^{1/2}d_w}\right|\nonumber\\
&=&4\left|(|z_0|-1)-\gamma\Re(z_0^{-2})
+\frac{3}{2(1+q_w)^{1/2}}\frac{\gamma}{d_w}\Re(z_0^{-3})\right|
\ .\end{eqnarray}
By comparing equations~(\ref{eqn:magc}) and (\ref{eqn:magw}),
we therefore establish the magnification correspondence 
(up to the order of $d_c^2\sim d_w^{-2}$) between
the close binary with $\hat Q=d_c^2 q_c (1+q_c)^{-2}$ and 
the wide binary with $\gamma=d_w^{-2} q_w (1+q_w)^{-1}$
when $\hat Q\simeq\gamma\ll 1$.

For the two PLANET ISIS models for \mb, 
we obtain $\hat Q = 3.40\times 10^{-3}$ for the close binary
solution $(d_c,q_c)=(0.134,0.340)$ and $\gamma=3.35\times 10^{-3}$
for the wide binary solution $(d_w,q_w)=(11.31,0.751)$. Hence,
the observed degeneracy between the two PLANET models (see
\S~\ref{sec:model} and Tables~\ref{tab:model} and \ref{tab:modelw})
is the clearest example of this type of correspondence
observed so far. While \citet{smc} reported that two different
binary lens models, one of a close binary and the other of a wide
binary, can fit the observed light curve of the (caustic-crossing)
binary lens event, \smc, \footnote{\citet{ob23} also found a similar
degeneracy while they modeled the light curve of \obc, but they could
discriminate the two models to better than $\Delta\chi^2\simeq 128$.}
one can infer from their figure~8, that the degeneracy between the two
models exists only for the specific light curves (which is essentially
a particular one dimensional slice of the magnification field over
the source plane) but not for the magnification field in the neighborhood
of the caustic as a whole. This is obvious from the relative rotation of
the two caustics. In addition, the source magnitudes for the two models
of \smc\ differ by $\sim 0.18\ \mbox{mag}$. By contrast, the difference
of the predicted $I_{\rm s}$ between the two models of \mb\ is only
$\sim 0.02\ \mbox{mag}$. In fact, we get 
$\hat Q=6.5 \times 10^{-2}$ (the close binary) and
$\gamma=1.8 \times 10^{-2}$ (the wide binary) for the two models
of \smc, and thus, although the degeneracy of the \smc\ light curve
is somehow related to the correspondence of $d\leftrightarrow d^{-1}$,
it cannot be completely explained simply by the argument in this appendix,
and it should
be investigated further for its origin. On the other hand, the degeneracy
of \mb\ is the first definitive observational case of the correspondence
between extreme separation binaries. This can be also seen in 
Figure~\ref{fig:caust}, which illustrates the similarity between the
magnification fields for the two models of \mb.

The very low $\Delta\chi^2$ between the two solutions despite
the excellent data implies that it is extraordinarily difficult
to break this degeneracy with photometric data. \citet{astro}
showed that for \smc\ the two solutions were also astrometrically
degenerate, at least for data streams lying within a few $t_{\rm E}$ of
the peak.  They argued that this astrometric degeneracy, like the
corresponding photometric degeneracy, was rooted in the lens equation.
However, as shown in this appendix, the correspondence between the
equations describing close- and wide-binaries is {\it purely local}
\citep[See also][]{d01}.
For example, there is a constant-offset term in equation~(\ref{eqn:crl}),
which does not give rise to any local photometric or astrometric effects,
but which must ``disappear'' at late times.  Hence there must be a
late-time
astrometric shift between the two solutions.   Such a shift was noted
explicitly by \citet{astro} for the case of \smc, and
these are likely to be a generic feature of close/wide corresponding
pairs of solutions.

\begin{figure}
\plotone{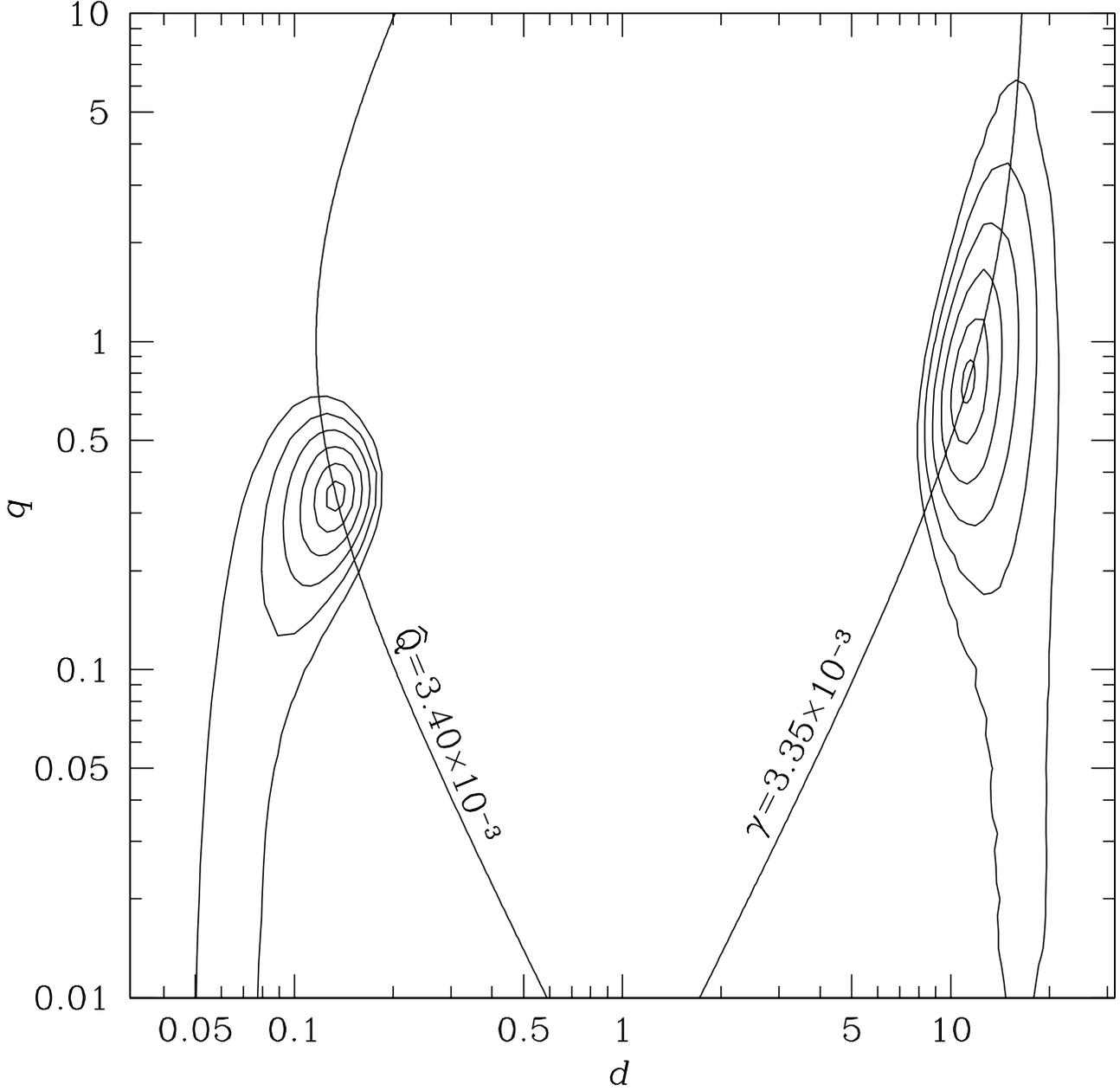}
\figcaption{
Contour plot of $\chi^2$-surface over ($d$,$q$) space based on
solutions for PLANET (ISIS) and MACHO data.
The binary separation $d$ is in units of the Einstein radius
of the combined mass, and the mass ratio $q$ is the ratio of
the farther component to the closer component to the source
trajectory (i.e., $q>1$ means that the source passes by the
secondary). Contours shown are of $\Delta\chi^2=1,4,9,16,25,36$
(with respect to the global minimum).
We find two well-isolated minima of $\chi^2$, one in the close-binary
region, ($d$,$q$)=(0.134,0.340), and the other in the wide binary-region,
($d$,$q$)=(11.31,0.751) with $\Delta\chi^2=0.6$ and the close binary
being the lower $\chi^2$ solution. Also drawn are
the curves of models with
the same $\hat Q$ as the best-fit close-binary model
and the same $\gamma$ as the best-fit wide-binary model
(see Appendix~\ref{sec:ap}).
\label{fig:chi2cont}
}\end{figure}

\begin{figure}
\plotone{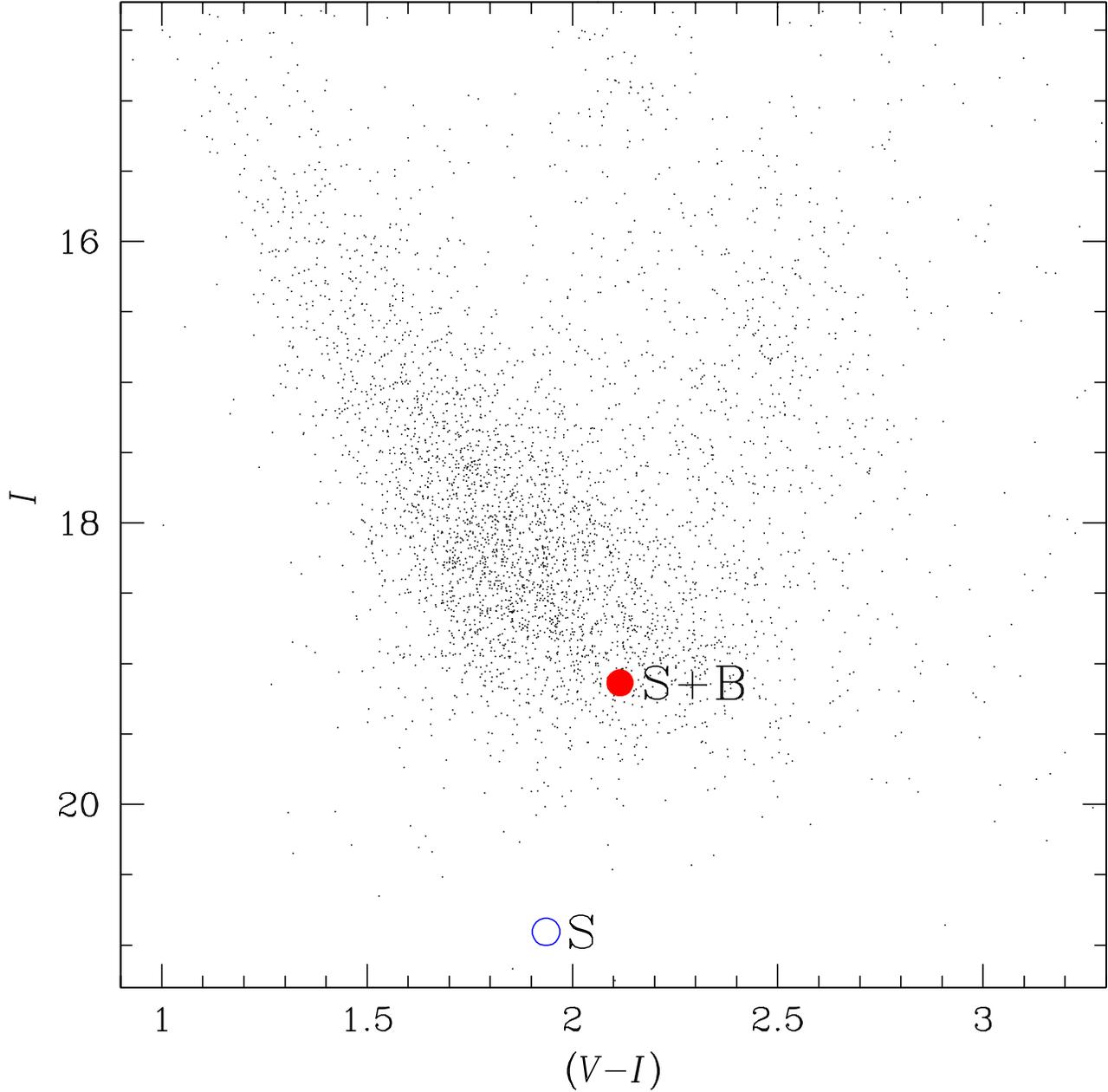}
\figcaption{
CMD of $3\arcmin\times3\arcmin$ field surrounding \mb\
($l = 17\arcdeg99$, $b= -1\fdg96$) derived from SAAO observations. 
Shown are the positions of the baseline ``star'' (S+B) and
the lensed source (S) in the absence of lensing (close-binary model).
The position of the lensed source in the wide-binary model differs
from this by substantially less than the size of the symbol.  The
errors are also smaller than the symbols.
The majority of stars in the CMD are most likely
turn-off stars seen at increasingly greater 
distances in the Galactic plane, and hence at correspondingly
greater reddenings.
\label{fig:cmd}
}\end{figure}

We also plot the lines of ($d$,$q$) pairs that have
the same shear $\gamma$ as the best-fit wide-binary model or
the same quadrupole moment $\hat Q$ as the best-fit close-binary model
on $\chi^2$-surface contour plot shown in Figure~\ref{fig:chi2cont}.
While the iso-shear line for wide binaries lies nearly parallel
to the direction of the principal conjugate near the best-fit
model, it is clear from the figure that the condition of
$\hat Q\simeq\gamma\ll 1$ alone does not define the observed
well-defined two-fold degeneracy, which involves the additional
correspondence between higher order terms beyond the quadrupole
moment ($\sim d_c^2$) and the pure shear ($\sim d_w^{-2}$).
Further comparison between equations~(\ref{eqn:magc}) and
(\ref{eqn:magw}) indicates that there exists a magnification
correspondence up to the order of $d_c^3\sim d_w^{-3}$ if the
condition $d_c(1-q_c)(1+q_c)^{-1}=d_w^{-1}(1+q_w)^{-1/2}$ is also
satisfied in addition to $\hat Q=\gamma\ll 1$.
We find that, for the two PLANET models,
$d_c(1-q_c)(1+q_c)^{-1}=6.6\times 10^{-2}$ (the close binary) and
$d_w^{-1}(1+q_w)^{-1/2}=6.7\times 10^{-2}$ (the wide binary). 
Hence, we conclude that, in fact, these two conditions,
\begin{equation}
d_c^2 d_w^2 (1+q_w) = \frac{q_w}{q_c} (1+q_c)^2
\,,\end{equation}
\begin{equation}
d_c d_w (1+q_w)^{1/2} = \frac{1+q_c}{1-q_c}
\,,\end{equation}
define a unique correspondence between two extreme separation
binaries.
We also note that the images (not shown) of the iso-$\Delta\chi^2$ contours
for the close binary models of \mb\ under the mapping defined by the
above two relations
follows extremely closely the corresponding iso-$\Delta\chi^2$ contour
for wide binary models, except for the difference of $\Delta\chi^2\simeq 0.6$
offset between two solutions.

\begin{deluxetable}{cccccc}
\tablecaption{
Photometric data of \mb
\label{tab:data}}
\tablehead{
\colhead{telescope}&
\colhead{filter}&
\colhead{\# of total points}&
\colhead{\# of points analyzed}&
\colhead{median seeing}&
\colhead{error rescaling factor}\\
&&&&(arcsec)&
}\startdata
\cutinhead{PLANET (DoPHOT)}
Elizabeth 1 m &$I$& 98& 88&1.81&1.19\\
\nodata       &$V$& 26& 24&1.79&1.02\\
  Canopus 1 m &$I$& 51& 42&2.71&1.43\\
     YALO 1 m &$I$&128&122&1.96&0.772\\
\cutinhead{PLANET (ISIS)}
Elizabeth 1 m &$I$&105& 98&1.65&1.35\\
\nodata       &$V$& 21& 20&1.96&1.83\\
  Canopus 1 m &$I$& 57& 38&2.59&1.75\\
     YALO 1 m &$I$&114&110&1.83&0.919\\
\cutinhead{MACHO}
Mt.Stromlo 50$''$ &$R_{\rm MACHO}$&90&90&\nodata&1.04\\
\nodata           &$V_{\rm MACHO}$&77&76&\nodata&
0.709\,\tablenotemark{a}\ /0.730\,\tablenotemark{b}
\enddata
\tablenotetext{a}
{with respect to PLANET/DoPHOT model}
\tablenotetext{b}
{with respect to PLANET/ISIS model}
\end{deluxetable}
\clearpage

\newpage
\begin{deluxetable}{ccc}
\tablecaption{
PLANET close-binary model of \mb
\label{tab:model}}\tablehead{
\colhead{parameters}&
\colhead{PLANET (ISIS) + MACHO}&
\colhead{PLANET (DoPHOT) + MACHO}
}\startdata
$d$&
$0.134\pm 0.009$&
$0.121\pm 0.009$\\
$q$&
$0.340\pm 0.041$&
$0.374\pm 0.058$\\
$t_{\rm E}$&
$163\pm 26\ \mbox{days}$&
$183\pm 32\ \mbox{days}$\\
$t_0$\,\tablenotemark{a,} \ \ \tablenotemark{b}&
$1393.9331\pm 0.0071$&
$1393.9309\pm 0.0075$\\
$u_0$\,\tablenotemark{a}&
$(8.6\pm 1.1)\times 10^{-3}$&
$(7.5\pm 1.3)\times 10^{-3}$\\
$\alpha$\,\tablenotemark{c}&
$294\fdg 99\pm 0\fdg 25$&
$294\fdg 45\pm 0\fdg 25$\\
$\chi^2$&
$412.13$&
$420.45$\\
dof&
412&
423
\enddata
\tablecomments
{We note that the fact that $\chi^2\sim\mbox{dof}$ results from
our rescaling of the photometric errorbars.
However, we use the same scaling factor here and for models described
in Table~\ref{tab:modelw} so the indistinguishability between two models
is not affected by this rescaling. The uncertainties for parameters
are ``1-$\sigma$ error bars'' determined by a quadratic fit of $\chi^2$ surface.}
\tablenotetext{a}
{the closest approach to the binary center of mass}
\tablenotetext{b}
{$\mbox{the Heliocentric Julian Date} - 2450000.$}
\tablenotetext{c}
{the binary center of mass lying on the right hand side of the moving source}
\end{deluxetable}

\begin{deluxetable}{ccc}
\tablecaption{
PLANET wide-binary model of \mb
\label{tab:modelw}}\tablehead{
\colhead{parameters}&
\colhead{PLANET (ISIS) + MACHO}&
\colhead{PLANET (DoPHOT) + MACHO}
}\startdata
$d$&
$11.31\pm 0.96$&
$12.79\pm 1.05$\\
$q$&
$0.751\pm 0.193$&
$0.917\pm 0.288$\\
$t_{\rm E}$\,\tablenotemark{a}&
$220\pm 37\ \mbox{days}$&
$253\pm 47\ \mbox{days}$\\
$t_0$\,\tablenotemark{b,} \ \ \tablenotemark{c}&
$1393.9113\pm 0.0072$&
$1393.9133\pm 0.0074$\\
$u_0$\,\tablenotemark{a,} \ \ \tablenotemark{b}&
$(6.5\pm 1.0)\times 10^{-3}$&
$(5.50\pm 0.92)\times 10^{-3}$\\
$\alpha$\,\tablenotemark{d}&
$295\fdg 46\pm 0\fdg 24$&
$294\fdg 83\pm 0\fdg 24$\\
$\chi^2$&
$412.74$&
$420.50$\\
dof&
412&
423
\enddata
\tablenotetext{a}
{with respect to the Einstein radius of the combined mass}
\tablenotetext{b}
{The closest approach to the caustic center.
See \S~\ref{sec:model} for the definition of the caustic center.}
\tablenotetext{c}
{$\mbox{the Heliocentric Julian Date} - 2450000.$}
\tablenotetext{d}
{the caustic center lying on the right hand side of the moving source}
\end{deluxetable}
\clearpage

\newpage
\clearpage

\newpage
\clearpage

\newpage
\clearpage

\newpage
\begin{figure}
\caption[f4.jpg]{
Grayscale plot of the difference of the normalized flux fields,
$2|F_w-F_c|/(F_{{\rm s},w}A_w + F_{{\rm s},c}A_c)$
of the two PLANET models around the caustic. 
The two fields are scaled and aligned so that the source
trajectories coincide with each other. The resulting shape
and the relative size of the caustics are remarkably close
to each other [c.f., fig.~8 of \citet{smc} and fig.~6 of \citet{ob23}].
Also drawn are the contours of zero difference (dotted line)
and 5\% difference (solid line). While the actual source trajectory
($y=0$) naturally traces well the zero-difference line,
the flux difference would also be extremely small for other
trajectories through the region, except very near the caustic.
\label{fig:caust}
}\end{figure}\clearpage

\end{document}